\documentstyle[11pt,newpasp,twoside,epsf]{article}
\def\ga{\mathrel{\raise.3ex\hbox{$>$\kern-.75em\lower1ex\hbox{$\sim$}}}}
\def\la{\mathrel{\raise.3ex\hbox{$<$\kern-.75em\lower1ex\hbox{$\sim$}}}}
\def\he#1{\hbox{${}^{#1}$He}}
\def\li#1{\hbox{${}^{#1}$Li}}
\def\be#1{\hbox{${}^{#1}$Be}}
\def\b#1#2{\hbox{${}^{#1#2}$B}}
\def\beq{\begin{equation}}
\def\eeq{\end{equation}}
\def\etal{{\it et al.~}}

\def\edcomment#1{\iffalse\marginpar{\raggedright\sl#1\/}\else\relax\fi}
\reversemarginpar

\begin{document}
\rightline{UMN--TH--1840/00}
\rightline{TPI--MINN--00/06}
\rightline{January 2000}  
\vskip -.31in
\title{The Evolution of \he4 and LiBeB}
 \author{Keith A. Olive}
\affil{Theoretical Physics Institute, School of
Physics and Astronomy,  University of Minnesota,
    Minneapolis MN 55455, USA}

\begin{abstract}
Our understanding of the evolution of $^4$He and $^7$Li
depends critically on the available data for these two elements at low metallicity.
In particular, the degree to which there is a slope in an abundance
vs metallicity regression can help determine the evolution of He, C, N, and O in
dwarf galaxies in the case of $^4$He, and cosmic-ray induced
nucleosynthesis of LiBeB in our own galaxy in the case of $^7$Li. Recent
data and their implications will be discussed.
\end{abstract}

\section{Introduction}

There is a relatively large set of data available on the big bang
nucleosynthesis (BBN) element isotopes of \he4 and \li7 at low metallicity. 
It is common, particularly in the case of \he4, to perform a linear
regression on the data with respect to some metallicity tracer such as O/H
or Fe/H. The intercept of such a regression can be directly related to
the primordial abundance of that isotope, and the slope of the regression
offers important clues to the nature of its chemical evolution.  
While one can not necessarily justify a {\em linear} relation from first
principles, generally due to the quality of the data at low metallicity, such
an approximation is acceptable.  In fact, using the currently available
\he4 and \li7 data\footnote{Using, for example, the data of Izatov and Thuan (1998)
on \he4 and Ryan, Norris, \& Beers (1999) on \li7.}, it is easy to show that a
linear regression is significantly better than a weighted mean, yet more complicated
fits using additional parameters generally do not yield a statistically
significant improvement in the fit. 

Our inferences of the primordial abundances and evolution of the light
elements are clearly tied to the quality of the data and our understanding of
the systematic uncertainties in the derived abundances.  Evolution is one the
effects which is responsible for systematic uncertainties.  In the case of
\he4, the environment of the HII system is  not pristine and includes
non-primordial \he4. In addition, the true elemental
abundances of \he4 may be clouded due to effects such as underlying stellar
absorption, collisional excitation, or flourecence. In the case of \li7, abundances
are contaminated by the non-primordial contribution of \li7 from galactic cosmic-ray
nucleosynthesis (GCRN) and uncertainties concerning the degree of stellar
depletion of \li7 in pop II, halo stars. 

\section{\he4}

The \he4 abundance has been
determined from observations  of HeII
$\rightarrow$ HeI recombination lines in a large sample of
extragalactic HII regions (Pagel \etal 1992; Skillman \& Kennicutt 1993;
Skillman \etal 1994; Izotov, Thuan,  \& Lipovetsky 1994,1997; Izotov \& Thuan
1998). Since \he4 is produced in stars along with heavier elements such as
Oxygen, it is expected that the primordial abundance of \he4 can be
determined from the intercept of the correlation between $Y$ and O/H, namely
$Y_p = Y({\rm O/H} \to 0)$.   A detailed analysis of the combined data
(Olive, Skillman, \& Steigman 1997; Fields \& Olive 1998) found an intercept
corresponding to a primordial abundance  
\beq
Y_p = 0.238 \pm 0.002 \pm 0.005
\label{he4}
\eeq
The first uncertainty is purely statistical and the second uncertainty is
an estimate of the systematic uncertainty in the primordial abundance
determination. The helium abundance used in this analysis was determined
using electron densities $n$ obtained from SII data. 
Izotov, Thuan,  \& Lipovetsky (1994,1997) and Izotov \& Thuan
(1998) proposed a method based on several He emission lines to
``self-consistently" determine the electron density.  This method yields a
higher primordial value
\beq
Y_p = 0.244 \pm 0.002 \pm 0.005
\label{he42}
\eeq

Our interpretation of the evolution of \he4 depends heavily on the slope of
the \he4 abundance with respect to a tracer element such as O/H and/or N/H. 
While models of chemical evolution tend to give relatively low slopes ($\Delta
Y/\Delta(O/H) \sim 20 - 60$), the He data based on SII densities gives
a much larger slope ($\Delta Y/\Delta(O/H) \sim 110 \pm 25$), whereas the
self-consistent method gives  ($\Delta Y/\Delta(O/H) \sim 47 \pm 26$).
The model calculations (Fields \& Olive 1998, and references therein) depend
crucially on the assumed yields of N in the AGB phase and on assumptions
concerning hot-bottom burning. Many of the models attempting to reproduce the
higher He slopes also rely on significant amounts of outflow in these dwarf
galaxies. 

As can be ascertained from the brief discussion above, the method of
analysis has a huge impact on both the determination of the primordial
\he4 abundance and the slope of the He vs O/H regression. Therefore, rather
than discuss specific chemical evolution models in detail here, I will
discuss some of the key sources of the uncertainties in the He abundance
determinations and prospects for improvement. 

The He abundance is always quoted relative to H, e.g., He line strengthes are
measured relative to 
$H\beta$.  The H data must first be corrected
for underlying absorption and reddening. 
Beginning with an observed line flux $F(\lambda)$, and an equivalent width
$W(\lambda)$, we can parameterize the correction for underlying stellar absorption as
\beq
X_A(\lambda) = F(\lambda) ({W(\lambda) + a) \over W(\lambda)})
\eeq
The parameter $a$ is expected to be relatively insensitive to 
wavelength.
A reddening correction is applied to determine the
intrinsic line intensity $I(\lambda)$ relative to $H\beta$
\beq
X_R(\lambda) = {I(\lambda) \over I(H\beta)} = {X_A(\lambda) \over X_A (H\beta)}
10^{f(\lambda) C(H\beta)}
\eeq
where $f(\lambda)$ represents an assumed universal reddening law and $C(H\beta$) is
the correction factor to be determined. By comparing $X_R(\lambda)$ to theoretical
values,
$X_T(\lambda)$, we determine the parameters
$a$ and $C(H\beta)$ self consistently (Olive \& Skillman, 2000), and run a Monte
Carlo over the input data to test the robustness of the solution and to
determine the systematic uncertainty associated with these corrections. 

In Figure 1 (from Olive \& Skillman 2000), I show the result of such a Monte-Carlo
based on synthetic data with an assumed correction of 2 \AA\ for underlying
absorption and a value for
$C(H\beta) = 0.1$.  The synthetic data were assumed to have an intrinsic 
2\% uncertainty. While the mean value of the Monte-Carlo results very
accurately reproduces the input parameters, the
spread in the values for $a$ and $C(H\beta)$ are generally a factor of 2 larger than
one would have derived from the direct solution due to the covarience in $a$ and
$C(H\beta)$. 

\begin{figure}[h] 
\epsfysize=6cm
\hskip 1in 
\epsfbox{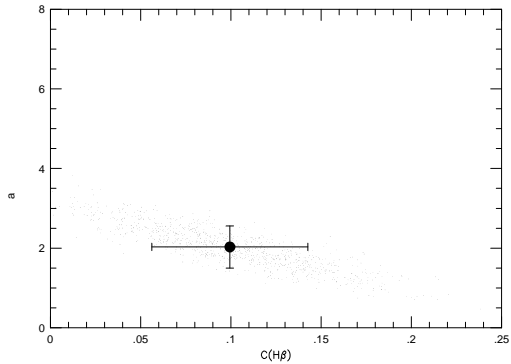}
\caption{A Monte Carlo determination of the underlying absorption $a$ (in \AA\ ), and
reddening  parameter $C(H\beta)$, based on synthetic data.}
\end{figure}

The uncertainties found for $H\beta$ must next be propagated into  the
analysis for
\he4, for which we follow an analogous procedure to that described above
(Olive \& Skillman 2000). We
again start with a set of observed quantities: line intensities
$I(\lambda)$ which include the reddening correction previously determined along with
its associated uncertainty which includes the uncertainties in $C(H\beta)$;
the equivalent width
$W(\lambda)$; and temperature $t$. The Helium line intensities are scaled to $H\beta$
and the singly ionized helium abundance is given by
\beq
y^+(\lambda) = {I(\lambda) \over I(H\beta)} {E(H\beta) \over
E(\lambda)} ({W(\lambda) + a'
\over W(\lambda)}) {1\over (1+\gamma)} {1\over f(\tau)}
\label{y+}
\eeq
where $E(\lambda)/E(H\beta)$ is the theoretical emissivity scaled to $H\beta$.
The expression (5) also contains a correction factor for underlying
stellar absorption, parameterized now by $a'$, a density dependent collisional
correction factor, $(1+\gamma)^{-1}$, and a flourecence correction which depends on
the optical depth $\tau$.  Thus $y^+$ implicitly depends on 3 unknowns, the
electron density, $n$, $a'$, and $\tau$. 

One can use 3-6 lines to determine the
weighted average helium abundance, $\bar y$.
From ${\bar y}$, we can calculate the $\chi^2$ deviation from the average,
and minimize $\chi^2$, to determine 
$n, a'$, and $\tau$. Uncertainties in the output parameters are also
determined. 

This procedure differs somewhat from that proposed by ITL, in that the
$\chi^2$ above is based on a straight weighted average, where as ITL
minimize the difference of a ratio of He abundances (to one wavelength, say
$\lambda$4471) to the theoretical ratio. When the reference line is particularly
sensitive to a systematic effect such as underlying stellar absorption, this
uncertainty propagates to all lines this way. In our case, the individual
uncertainties in the line strengths are kept separate.

Finally, as in the case for the hydrogen lines, we have performed a Monte-Carlo
simulation of the data to test the robustness of the solution for $n,a'$, 
and $\tau$ (Olive \& Skillman 2000). In Figure 2, I show the result of a single
case based on the data of  Izatov and Thuan (1998) for SBS1159+545.
Here, the helium abundance and density solutions are displayed.
The vertical and horizontal lines show the position of the IT solution.
The circle shows the position of the our solution to the minimization, and
the square shows the position of the mean of the Monte-Carlo distribution. 
The spread shown here is significantly greater than the uncertainty quoted by IT.

\begin{figure}[h] 
\epsfysize=6cm
\hskip 1in 
\epsfbox{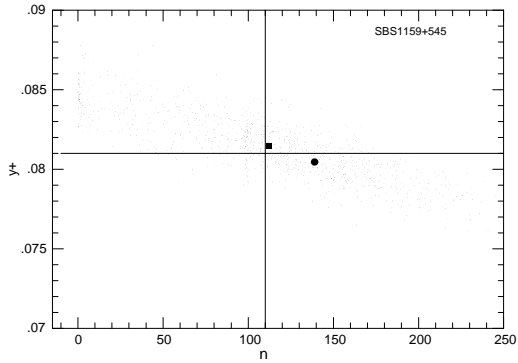}
\caption{A Monte Carlo determination of the helium abundance and electron density
(in cm$^{-3}$). Solutions for $a^\prime$ and $\tau$ are not shown here. }
\end{figure}

\section{\li7}

The  population II abundance of \li7 has been determined by observations of
over 100 hot, halo stars, and is found to have a very
nearly  uniform abundance (Spite \& Spite 1982). For
stars with a surface temperature $T > 5500$~K
and a metallicity less than about
1/20th solar, the  abundances show little or no dispersion beyond that which
is consistent with the errors of individual measurements. The Li data from
Bonifacio \& Molaro (1997)
indicate a mean \li7 abundance of 
\beq
{\rm Li/H = (1.6 \pm 0.1 ) \times 10^{-10}}
\label{li}
\eeq
The small error is statistical and is due to the large number of stars
in which \li7 has been observed. 

There is, however, an important source of systematic error due to the
possibility that Li has been depleted in these stars, though the lack of
dispersion in the Li data limits the amount of depletion. In fact, as
discussed by Sean Ryan (these proceedings, and Ryan, Norris, \& Beers, 1999, hereafter
RNB) a small observed slope in Li vs Fe
and the tiny dispersion about that correlation indicates that
depletion is negligible in these stars. Furthermore, the slope may indicate a
lower abundance of Li than that in (6).

For reference, the weighted mean of the \li7 abundance in the RNB sample 
is [Li] = 2.12 ([Li] = $\log$ \li7/H + 12).  It is common to test for the presence
of a slope in the Li data by fitting a regression of the form [Li] = $\alpha +
\beta$ [Fe/H]. The RNB data indicate a rather large slope, $\beta = 0.07 - 0.16$ and
hence a downward shift in the ``primordial" lithium abundance $\Delta$[Li] = $- 0.20
- - 0.09$.  Models of galactic evolution which predict a small slope for [Li] vs.
[Fe/H], can produce a value for $\beta$ in the range 0.04 -- 0.07 (Ryan \etal 2000). 
Of course, if we would like to extract the primordial \li7 abundance, we must examine the
linear (rather than log) regressions. 
For Li/H = $a^\prime + b^\prime$Fe/Fe$_\odot$, we find $a^\prime = 1 - 1.2 \times
10^{-10}$ and $b^\prime = 40 - 120 \times
10^{-10}$.  A similar result is found fitting Li vs O. 
Overall, when the regression based on the data and other systematic effects are
taken into account a best value for Li/H was found to be (Ryan \etal 2000)
\beq
{\rm Li/H = 1.23  \times 10^{-10}}
\label{li2}
\eeq
with a plausible range between 0.9 -- 1.9 $\times 10^{-10}$. 

\begin{figure}[h]
\epsfysize=6cm
\hskip 1in 
\epsfbox{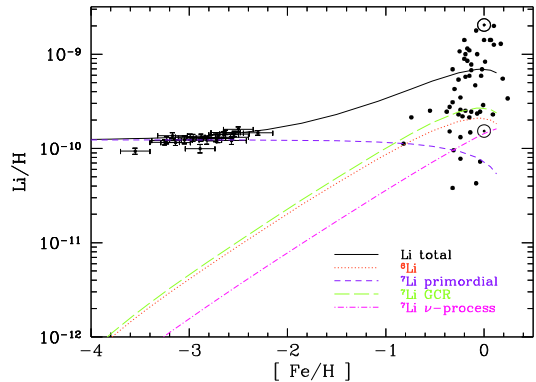}
\caption{Contributions to the total predicted lithium abundance from the adopted GCE
model of Fields \& Olive (1999a,b), compared with low metallicity 
stars (RNB) and a sample of high metallicity stars.
The solid curve is the sum of all  components.  }
\end{figure}

Figure 3 shows the different Li
components for a model with (\li7/H)$_p = 1.23 \times 10^{-10}$. 
The linear slope produced by the model is  $b' = 65\times 10^{-10}$,
and is independent of the input primordial value (unlike the log slope given above).
The model (discussed in detail in Fields \& Olive, 1999a,b and Fields \etal 2000) includes
in addition to primordial \li7, lithium produced in galactic cosmic ray nucleosynthesis, 
(primarily $\alpha + \alpha$ fusion) in addition to \li7 produced by the $\nu$-process
during type II supernovae. As one can see, these processes are not sufficient to 
reproduce the population I abundance of \li7, and additional production sources are
needed (see e.g. Matteucchi, these proceedings).

\section{Concordance}

Bearing in mind the degree of uncertainty in the derived primordial abundances, one can
test the concordance of \he4 and \li7 with the prediction of BBN.  This is best
summarized in a comparison of likelihood functions as a function of the one free
parameter of BBN, namely the baryon-to-photon ratio $\eta$. By combining the theoretical
predictions (and its uncertainties) with the observationally determined abundances
discussed above, we can produce individual likelihood functions (Fields \etal 1996) which
are shown in Figure 4. A range of primordial \li7 values are chosen based on the
analysis in Ryan \etal (2000). The double peaked nature of the \li7
likelihood functions is due to the presence of a minimum in the
 predicted lithium abundance in the expected range for $\eta$.  For a given observed
value of \li7, there are two likely values of $\eta$. As the lithium abundance is
lowered, one tends toward the minimum of the BBN prediction, and the two peaks merge. 
Also shown are both values of the primordial \he4 abundances discussed above. 
As one can see, at this level there is clearly concordance between \he4, \li7 and BBN.

\begin{figure}[h]
\epsfysize=6cm
\hskip 0.5in 
\epsfbox{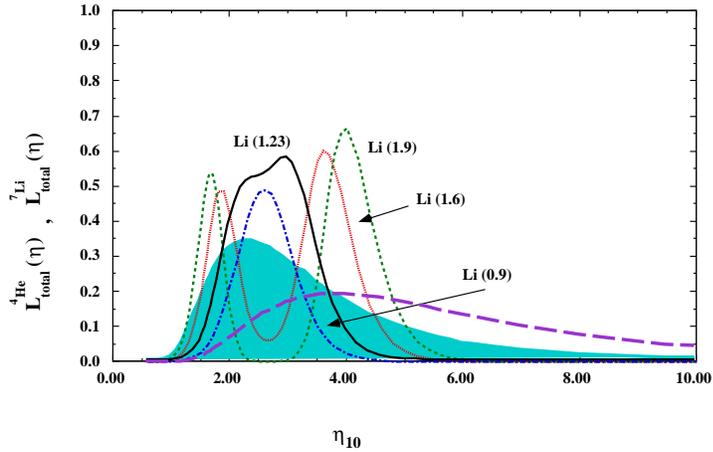}
\caption{Likelihood distributions for four values of primordial $^7$Li/H
($10^{10}\times$ \li7 = 1.9 ({\it dashed}), 1.6 ({\it dotted}), 1.23 ({\it
solid}), and 0.9 ({\it dash-dotted})), and for $^4$He ({\it shaded}) for which we
adopt $Y_p = 0.238\pm 0.002\pm 0.005$ (Eq. (1)). Also shown by the long dashed curve
is the likelihood function based on the \he4 abundance from Eq. (2). }
\end{figure}

\section {LiBeB}

The production of \li7 by galactic cosmic-ray nucleosynthesis shown in Figure 3, is
accompanied by the production of the heavier intermediate elements Be and B (Reeves,
Fowler, \& Hoyle 1970; Meneguzzi, Audouze, \& Reeves 1971). 
Standard GCRN is dominated by interactions originating from
accelerated protons and $\alpha$'s on CNO in the ISM, and predicts that BeB should be 
``secondary'' versus the spallation targets, giving $\be{} \propto {\rm O}^2$ 
(Vangioni-Flam, Cass\'e, Audouze, \& Oberto 1990).
However, this simple model was challenged by the
observations of BeB abundances in Pop II stars,
and particularly the BeB trends versus metallicity.
Measurements showed that both
Be and B vary roughly {\em linearly} with Fe,
a so-called ``primary'' scaling.
If O and Fe are co-produced (i.e., if O/Fe is constant)
then the data clearly contradicts the canonical theory, i.e. BeB production via
standard GCR's.

These observations led to the creation of many new models of cosmic-ray nucleosynthesis
(Cass\'e, M., Lehoucq, R., \& Vangioni-Flam, E. 1995; Vangioni-Flam \etal 1996; Ramaty
\etal 1997, 1999) which include a dominant primary component of BeB. Such models were
discussed here by Cass\'e, Parizot and Ramaty.  

As was discussed here by Deliyannis and Israelian, there is growing evidence that the
O/Fe ratio is {\em not} constant at low metallicity (Israelian, Garc\'{\i}a-L\'{o}pez,
\& Rebolo 1998;  Boesgaard \etal 1999a), but rather increases towards low
metallicity. This trend offers
another solution 
to resolve discrepancy between the observed BeB
abundances as a function of metallicity and the predicted secondary trend
of GCR spallation (Fields \& Olive 1998). As noted above, standard GCR nucleosynthesis
predicts $\be{}\propto {\rm O}^2$, while
observations show $\be{} \sim {\rm Fe}$, roughly; 
these two trends can be consistent if O/Fe is not constant
in Pop II.   A combination of standard GCR nucleosynthesis, and
$\nu$-process production of \b11 is consistent with current data. 

The determination of abundances from
raw stellar spectra requires stellar
atmosphere models.
The atmospheric models require key input parameters, notably
the effective temperature $T_{\rm eff}$ and surface gravity $g$,
and assumptions regarding the applicability of local thermodynamic equilibrium (LTE).
Unfortunately, there is no
standard set of stellar parameters for the halo stars of interest. 
In practice, different groups derive abundances via
different procedures, which
give similar results but retain systematic differences.
The systematic differences in the data can in fact obscure the  BeB-OFe
trends one seeks. Thus, to derive meaningful BeB fits,
one must systematically and consistently
present abundances derived under the same assumptions
and parameters for stellar atmospheres.

It is not possible to overly stress the importance of reliable stellar data.
The Balmer-line data appear to be self-consistent, and are probably the
most reliable.  However, because other scales such as those based on the
IRFM are commonly used, I would like to point out that there are
significant differences in the reported data.  To illustrate the point consider for
example the case of the star BD $3^\circ$ 740. {}From Axer \etal (1994), whose data is
based on the Balmer line method we find this star to have  ($T_{\rm eff}, \ln g$,
[Fe/H]) = (6264, 3.72, -2.36). The beryllium and oxygen abundances for this star was
reported by Boesgaard \etal (1999b) and (1999a).  When adjusted for
these stellar parameters, we find [Be/H] = -13.36, and [O/H] = -1.74.
In contrast, the stellar parameters from Alonso \etal  (1996a) based on the IRFM (IRFM1)
are (6110,3.73,-2.01) with corresponding Be and O abundances of
-13.44 and -2.05.  Garcia-Lopez \etal 1998 use a calibrated
IRFM (IRFM2) based on Alonso \etal (1996b) and take (6295,4.00, -3.00).
For these choices, we have [Be/H] = -13.24 and [O/H] = -1.90.
Notice the extremely large range in assumed metallicities and the
difference in the two so-called IRFM temperatures.  While this may not be
a typical example of the difference in stellar parameters,  it is
differences such as this (and this star is not unique) that accounts for
the difference in our results and the implications we must draw from them.
Uncertainties in [Fe/H] in particular, make modeling extremely difficult.
This is especially true of one attempts to model the correlations of BeBO with respect to
Fe/H.

Below, we will present results based for the available BeBOFe data based
on three methods of analysis. We will refer to these as the Balmer line
data and the IRFM1,2 data.  Complete results of this analysis can be found in 
Fields \etal (2000). There are a total of 36 stars with low metallicity OH data. 
Of these, roughly 2/3 have available data using one of the systematic methods described
(Balmer, IRFM1, IRFM2). In each case, one finds a significant slope for [O/Fe] vs [Fe/H]
ranging from -0.32 to -0.51.

Of key importance to the modeling of the BeB evolution is the determination of 
a primary or secondary source for the BeB isotopes. Primary vs secondary is typically 
ascertained by fitting the BeB data versus a tracer element.  Historically, Fe/H was
used even though the actual production of LiBeB is independent of [Fe/H]. This is
justifiable so long as [O/Fe] is constant. As one can see in the tables below, the data
seem to indicate that Be is mostly primary with respect to Fe, and secondary with
respect to O/H. (IRFM2 should be considered suspect as the derived parameters were
obtained outside the limits of validity of the calibration.)  This is what one would
expect if [O/Fe] is not constant as the OH data now indicate.  B on the other hand shows 
primary evolution with respect to [O/H] and an even flatter evolution with respect to 
Fe/H.  This too is expected if the $\nu$-process plays a significant role in the
production of \b11. 

\begin{table}[htb]
\caption{Slopes for Be versus Fe and O.}
\begin{tabular}{ccccccc}
\hline\hline
  method & number & tracer & slope &  number & tracer & slope\\
\hline
 Balmer & 22 &  Fe &  $1.39 \pm 0.16$ & 19 &  O  &  $1.78 \pm 0.19$  \\
IRFM1 & 22 &   Fe &   $1.23 \pm 0.14$ & 21 &   O  &   $1.83 \pm 0.19$\\
IRFM2 & 18 &   Fe &   $1.18 \pm 0.11$ & 18 &   O  &   $1.36 \pm 0.09$\\
\hline\hline
\end{tabular}
\label{bedata}
\end{table}

\begin{table}[htb]
\caption{Slopes for B versus Fe and O.}
\begin{tabular}{ccccccc}
\hline\hline
method & number & tracer & slope &  number & tracer & slope \\
\hline
Balmer & 11 &  Fe &  $0.78 \pm 0.22$ & 10 &  O  &  $1.23 \pm 0.32$\\
IRFM1  & 9 &   Fe &   $0.73 \pm 0.19$ & 9 &   O  &   $0.98 \pm 0.28$\\
IRFM2 & 11 &   Fe &   $0.72 \pm 0.14$ & 11 &   O  &   $1.02 \pm 0.16$\\
\hline\hline
\end{tabular}
\label{bdata}
\end{table}

\begin{figure}[h]
\epsfysize=10cm
\hskip 1in 
\epsfbox{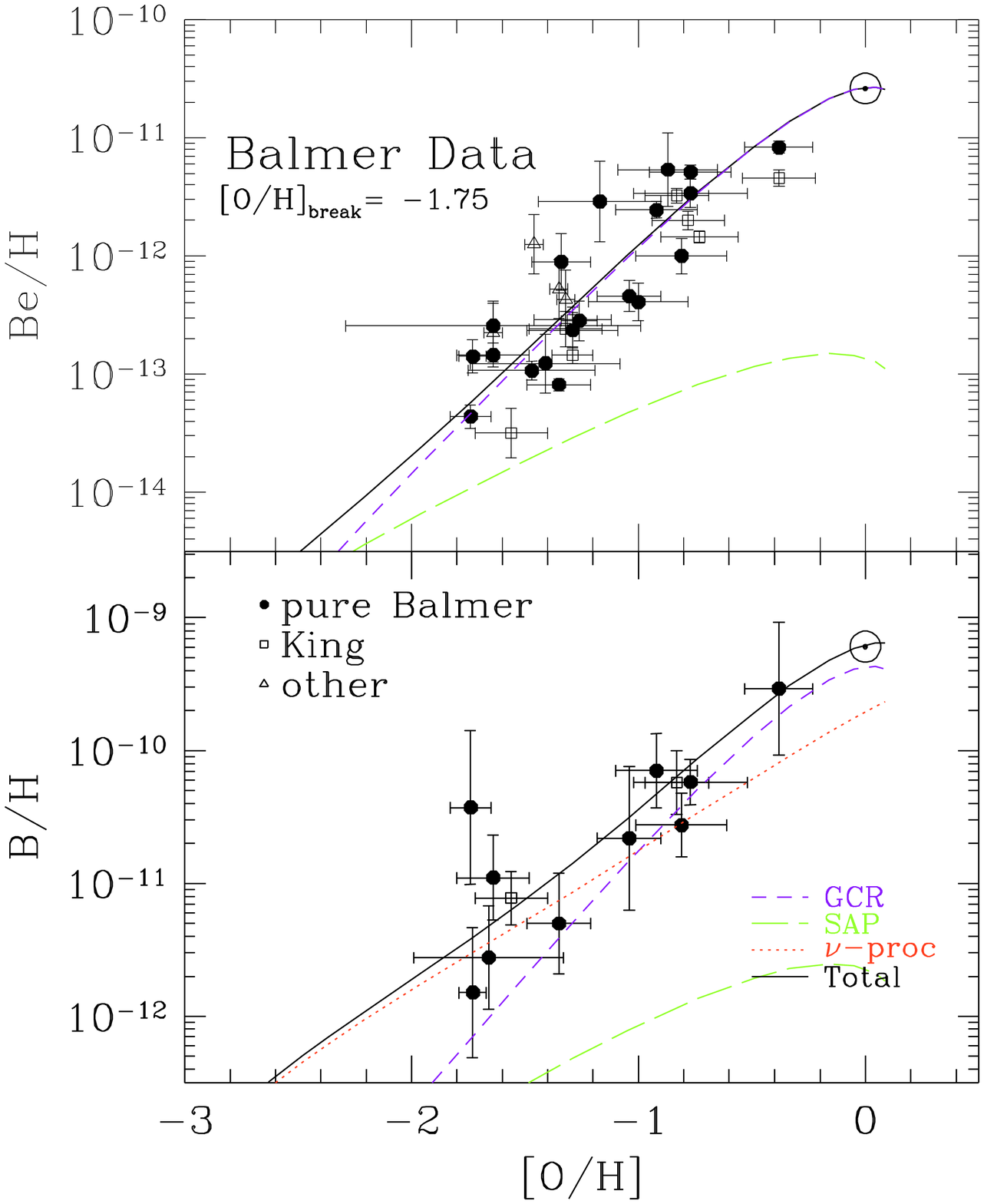}
\caption{Be {\it vs} O ({\it top panel}) and B {\it vs} O ({\it bottom panel}).  
Data shown are the Balmer
points, which are found to have a break
point as indicated.  Models are adjusted to have the
break point and O/Fe slope of these data.
}
\end{figure}

The models for primary and secondary
production of BeB are all physical.  What is unclear however, is which is dominant
over the history of the Galaxy and at what epoch.  If both mechanisms are operative,
it is reasonably certain that primary mechanisms should dominate in the early Galaxy
and that secondary mechanisms should dominate later. The cross-over or break point
can be determined from the data (in principle) by fitting to both linear and
quadratic components,
\beq
\frac{A}{\rm H}  = \left( \frac{A}{\rm H} \right)_\odot
        \left[ \alpha_1 \frac{\rm O/H}{({\rm O/H})_\odot}
        + \alpha_2 \left(\frac{\rm O/H}{({\rm O/H})_\odot}\right)^2 \right]
\eeq
for $A \in {\rm BeB}$.
The resulting coefficients and break points for Be are found in Table 3.
As one can see, for Balmer and IRFM1, the break point occurs at low [O/H], indicating
that most of the evolution in the observed data has been secondary.  
To fully resolve this issue, a larger and systematic data set is required. 

\begin{table}[htb]
\caption{Break points for Be versus O.}
\begin{tabular}{ccccc}
\hline\hline
 number & method & $\alpha_1$ & $\alpha_2$ & [O/H]$_{\rm eq}$ \\
\hline
19 & Balmer &  $0.042 \pm 0.003$ &  $2.30 \pm 0.70$ & -1.75 \\
21 & IRFM1 &  $0.034 \pm 0.034$ &  $2.14 \pm 0.53$ & -1.79 \\
18 & IRFM2 &  $0.111 \pm 0.031$ &  $2.57 \pm 0.76$ & -1.37 \\
\hline\hline
\end{tabular}
\label{tab:break}
\end{table}

Finally, in Figure 5 (from Fields \etal 2000), the evolution of BeB with respect to
O/H is shown in a simple closed box model of chemical evolution.  In addition to
standard GCR nucleosynthesis, a primary component based on the superbubble
accelerated particle spectrum of Bykov (1999) is included along with the
neutrino-process for
\b11 (Fields \etal 2000).  The secondary GCR cosmic-ray flux is normalized by the
solar abundance of Be and is consisitent with the present cosmic-ray flux scaled by
the star formation rate. The break point (from Table 3) determined the relative
scaling of the primary component to the secondary one, and the neutrino process is
scaled to the solar
\b11/\b10 ratio.

\acknowledgements
I would like to thank my collaborators T. Beers, M. Cass\'e, B. Fields, J. Norris, S.
Ryan, E. Skillman, and E. Vangioni-Flam whose work has been summarized here. 
This work was supported in part by
DoE grant DE-FG02-94ER-40823 at the University of Minnesota.

\end{document}